\documentclass[useAMS,usenatbib]{mn2e}
\usepackage{graphicx,amssymb,color}
\usepackage[normalem]{ulem}
\newcommand{\rev}{ }

\title[High-resolution post-MS MMR portraits]
{High-resolution resonant portraits of a single-planet white dwarf system}

\author[]{Dimitri Veras$^{1,2,3}$\thanks{E-mail: dimitri.veras@colorado.edu},
Nikolaos Georgakarakos$^{4,5}$, Ian Dobbs-Dixon$^{4,5,6}$
\\
$^{1}$Centre for Exoplanets and Habitability, University of Warwick, Coventry CV4 7AL, UK
\\
$^{2}$Centre for Space Domain Awareness, University of Warwick, Coventry CV4 7AL, UK
\\
$^{3}$Department of Physics, University of Warwick, Coventry CV4 7AL, UK
\\
$^{4}$Division of Science, New York University Abu Dhabi, PO Box 129188, Saadiyat Island, Abu Dhabi, United Arab Emirates
\\
$^{5}$Center for Astro, Particle and Planetary Physics (CAP3), New York University Abu Dhabi, PO Box 129188, Saadiyat Island, 
\\
Abu Dhabi, United Arab Emirates
\\
$^{6}$Center for Space Science, NYUAD Institute, New York University Abu Dhabi, PO Box 129188, Abu Dhabi, United Arab Emirates
}

\pubyear{2023}

\begin{document}
\label{firstpage}
\pagerange{\pageref{firstpage}--\pageref{lastpage}}
\maketitle

\begin{abstract}
The dynamical excitation of asteroids due to mean motion resonant interactions with planets is enhanced when their parent star leaves the main sequence. However, numerical investigation of resonant outcomes within post-main-sequence simulations is computationally expensive, limiting the extent to which detailed resonant analyses have been performed. Here, we combine the use of a high-performance computer cluster and the general semianalytical libration width formulation of \cite{galetal2021} in order to quantify resonant stability, strength and variation instigated by stellar evolution for a single-planet system containing asteroids on both crossing and non-crossing orbits. We find that resonant instability can be accurately bound with only main-sequence values by computing a maximum libration width as a function of asteroid longitude of pericentre. We also quantify the relative efficiency of mean motion resonances of different orders to stabilize versus destabilize asteroid orbits during both the giant branch and white dwarf phases. The $4$:$1$, $3$:$1$ and $2$:$1$ resonances represent efficient polluters of white dwarfs, {\rev and} even {\rev when} in the orbit-crossing regime, both the $4$:$3$ and $3$:$2$ resonances can retain small reservoirs of asteroids in stable orbits throughout giant branch and white dwarf evolution. This investigation represents a preliminary step in characterising how simplified extrasolar Kirkwood gap structures evolve beyond the main-sequence.   
\end{abstract}


\section{Introduction}

The dynamics of post-main-sequence planetary systems are rich. After remaining relatively static for potentially billions of years along the main-sequence, planetary and small body architectures and reservoirs experience significant physical and orbital changes while and after their parent star convulses during its death throes \citep{veras2016}. 

The end result of many of these changes are manifest in the growing body of observations of planetary remnants close to and within the photospheres of white dwarfs. In fact, dedicated surveys suggest that over a quarter of all Milky Way white dwarfs are currently accreting planetary material \citep{zucetal2003,zucetal2010,koeetal2014}, with compositions that vary from rocky, terrestrial-like \citep{juryou2014,doyetal2019} to volatile-rich \citep{xuetal2017} and potentially consistent with primitive planetesimals \citep{xuetal2013,curetal2022}, asteroids or meteorites \citep{ganetal2012,wiletal2015,melduf2017,swaetal2019,bucetal2022}, comets \citep{faretal2013,radetal2015,hosetal2020} or moons \citep{doyetal2021,kleetal2021}.

The accretion of planetary matter onto white dwarfs -- a process which is commonly known as ``white dwarf pollution" -- has been observed directly \citep{cunetal2022} and is thought to almost always \citep{rocetal2015,bonetal2017} originate from Solar radii-scale circumstellar discs. These discs probably represent the debris of broken-up minor planets, and over 60 of these discs have now been observed \citep[e.g.][]{farihi2016,manetal2020}. Their formation has been described dynamically \citep{jura2003,debetal2012,veretal2014a,malper2020b,verkur2020,lietal2021,broetal2022}, and the nature of their progenitor bodies reflects the chemical composition seen in the photospheres of their parent white dwarfs.

This variation in composition and potential progenitor planetary bodies suggests that the dynamical mechanisms which so readily transport planetary material to the close vicinity of white dwarfs are system-specific. Nevertheless, a mechanism which is thought to be common is a sequence of gravitational perturbations generated by a surviving planet and imposed upon a minor body that creates a pathway for the latter to reach the Roche radius of (disruption distance from) the white dwarf. 

Mounting investigations of these pathways\footnote{See Fig. 6 of \cite{veras2021} for a full reference listing of publications as a function of planetary architecture.} reveal that they are more likely to be traversed with a  greater number of perturbers in the system, whether these be additional planets \citep{veretal2016,maletal2021,ocoetal2022} or additional stellar companions \citep{bonver2015,hampor2016,petmun2017,steetal2017,veretal2017}. However, in a significant fraction of systems, only one planet may survive. In fact, all known planets orbiting white dwarfs are in single-planet systems \citep{thoetal1993,sigetal2003,luhetal2011,ganetal2019,vanetal2020,gaietal2022}, including the notable case of a roughly Jovian analogue with respect to both planet-star separation and mass \citep{blaetal2021}.

In single-planet, single-star systems, the dynamical pathways for an asteroid or comet to reach the white dwarf are restricted. In the absence of external forces, a planet must be on an eccentric orbit in order to perturb an asteroid into a white dwarf \citep{antver2016}. The external force provided by the star as it traverses the giant branch phases and loses mass can change stability boundaries \citep{debsig2002,veretal2013a,musetal2014,vergan2015} and thereby allow minor bodies to reach the star \citep{bonetal2011,frehan2014,veretal2021}, but not always \citep{veretal2020}.  

One way to increase a minor body's eccentricity to a sufficiently high value to reach the white dwarf's Roche radius is for the minor body to be trapped in a mean-motion resonance with the planet. Although resonances often appear in the post-main-sequence planetary literature, only a few such investigations have actually focussed on the dynamics of resonance. \cite{voyetal2013} semi-analytically modelled how stellar mass loss changes the resonant equations of motion, and used the the maximal Lyapunov characteristic number as a chaos indicator to explore relevant regions of phase space. \cite{smaetal2018,smaetal2021} focussed on the efficiency of secular resonances induced in multi-planet systems after one planet was engulfed. Other studies \citep{debetal2012,caihey2017,veretal2018,antver2019,ronetal2020,veretal2021,verhin2021,lietal2022} have included helpful individual plots that demonstrate the reach of particular mean-motion resonances in post-main-sequence planetary systems.

In this paper, we provide a focussed and detailed numerical analysis of post-main-sequence mean-motion resonances in two similar one-planet systems, each containing $10^5$ asteroids and differing only in planetary mass. For perspective, this number of asteroids is an order-of-magnitude greater than the number of asteroids we used in the simulations of \cite{veretal2021}, and all {\rev of} the simulations here combine stellar and planetary evolution self-consistently {\rev by} using the slow but accurate code from \cite{musetal2018}. Here, we combine this computationally expensive suite of numerical integrations of these asteroids with an analysis enabled by employing the general semianalytical resonant characterization code of \cite{galetal2021}. 

A simplified version of this characterization code has been used in a suite of previous papers \citep{gallardo2006,gallardo2019,gallardo2020} and is particularly valuable because it is neither restricted to a particular resonant order nor subject to convergence failures of some disturbing function expansions. The code instead numerically calculates the disturbing function given a complete set of orbital elements and masses, and generates several useful outputs, the most relevant of which for this study is libration width. Such facility is enabled by assuming that the longitudes of pericentre and longitudes of ascending node of both planet and asteroid are fixed throughout the resonant motion (but remain selectable as initial conditions).

Our goals here are four-fold: (i) to provide a detailed resonant map for an evolved planetary architecture covering a wide range of resonant orders, (ii) to determine an efficient way to characterize these resonances, (iii) to determine which resonances can be efficient at perturbing asteroids towards the white dwarf, for this given architecture, and (iv) to determine which resonances can be efficient at ejecting asteroids from the system. This last point is particularly relevant to assess the likely origins of small interstellar free-floaters \citep{moromartin2022}, of which we might see an increasing number passing through the solar system with the {\it Vera C. Rubin Observatory}.

This paper is organized as follows. In Section 2, we describe our simulation setup. We analyze these results in Section 3, and then summarise in Section 4.

\section{Simulation setup}

Because our objective is to analyze resonant outcomes and constraints for a single-planet system in detail, we carefully chose our simulation initial conditions such that they would generate substantial statistics while still remaining physical and plausible. We were guided by the following considerations:

\begin{enumerate}

\item Both asteroids and the planet are subject to destruction from the parent star as it transitions from a main-sequence to a giant branch star. The increase in the star's size can tidally drag inward and engulf objects out to distances of several au \citep{kunetal2011,musvil2012,norspi2013,viletal2014,madetal2016,ronetal2020}, and the increase in the star's luminosity could break up asteroids out to tens of au \citep{veretal2014b,versch2020} and drag debris both inwards and outwards \citep{veretal2015a,veretal2019,feretal2022}. Hence, there is a minimum surviving distance for both the planet and asteroids, and asteroids could occupy a wide swathe of distance space. 

\item When the planet mass is too small, the parameter space widths of mean motion resonances with asteroids fail to exceed the resolution achieved in numerical simulations. In other words, in this case, too many asteroids would need to be sampled to clearly detect the resonant signature. On distance scales of au, this critical mass is comparable to that of a Super-Earth or mini-Neptune \citep{veretal2021}. In general then, at the present time, only giant planets produce sufficiently high-resolution signatures in simulations of evolved planetary systems on au-scales for our purposes.

\item Within the solar system, we see resonances between objects which are both on osculating crossing orbits (e.g. Neptune and Pluto) and osculating non-crossing orbits (e.g. some Hecuba gap asteroids). We also detect much higher order resonant signatures \citep[e.g. Fig. 20 of][]{deretal2021} than is currently achievable in observations of extrasolar planetary systems, although such signatures undoubtedly exist. Hence, the more types of mean-motion resonances we can sample, the better that we can reflect reality.

\item A planet on a circular orbit very rarely, if ever, can perturb an asteroid close to its parent star \citep{antver2016,veretal2021}. Hence, a planet on an eccentric orbit would need to be adopted in order to sample white dwarf pollution. Further, planets in evolved planetary systems which would survive to the white dwarf phase are rarely on exactly circular orbits \citep{gruetal2018,gruetal2022}, and can only be circularized through subsequent gravitational perturbations and tidal effects by other larger bodies in the same system \citep{verful2019,verful2020,munpet2020,ocolai2020,steetal2021} or migration within common envelopes \citep{chaetal2021,lagetal2021,szoetal2022,yaretal2022}.

\end{enumerate} 

These considerations have led us to choose a planetary architecture similar to one used in \cite{veretal2021}, where, at the end of the main sequence, the planet's semimajor axis is $a_{\rm pl} = 10.5$ {\rm au} and its eccentricity is $e_{\rm pl} = 0.2$. Asteroids are interior to the planet and fill the $(a_{\rm ast}, e_{\rm ast})$ parameter space at high resolution, subject to the restriction that their orbital pericentres exceed 2 au and $a_{\rm ast} = 4-9$ au. 

These choices allowed us to sample a variety of mean-motion commensurabilities: three of first-order ($2$:$1$, $3$:$2$, $4$:$3$), four of second-order ($3$:$1$, $5$:$3$, $7$:$5$, $9$:$7$), four of third-order ($4$:$1$, $5$:$2$, $7$:$4$, $8$:$5$), and numerous ones of slightly higher orders. Our choices also allowed us to sample crossing and non-crossing orbits.

The specifics of our simulation setup are as follows. We treated the asteroids as test particles. Hence, the planet's orbit was not affected by the asteroids, and we patched together many serial simulations to create a single high-resolution resonant portrait. We did so in order to create two composite simulations, each containing a total of $10^5$ test particles. In one composite, the planet's mass was a Saturn mass, and in the other, the planet's mass was a Jupiter mass.

For the other orbital elements of the objects, we set the planet's inclination $i_{\rm pl}$, argument of pericentre $\omega_{\rm pl}$ and longitude of ascending node $\Omega_{\rm pl}$ all to $0^{\circ}$. During stellar evolution, as the star loses mass, both $i_{\rm pl}$ and $\Omega_{\rm pl}$ remain fixed, but $\omega_{\rm pl}$ does vary \citep{omarov1962,hadjidemetriou1963,veretal2011}, assuming isotropic mass loss \citep{veretal2013b,doskal2016a,doskal2016b}. For the asteroids, we randomly sampled them from uniform distributions of $a_{\rm ast} = 3-9$~au and $e_{\rm ast}$, subject to their orbital pericentre $q_{\rm ast} \ge 2$ au. We randomly sampled $i_{\rm ast}$ values from uniform distributions from $0^{\circ}$ to $5^{\circ}$, and randomly chose $\omega_{\rm ast}$ and $\Omega_{\rm ast}$ from uniform distributions from $0^{\circ}$ to $360^{\circ}$.

The host star began as a $M_{\star}^{\rm MS} = 2.0M_{\odot}$ main-sequence star and evolved into a $M_{\star}^{\rm WD} = 0.637M_{\odot}$ white dwarf, according to the prescription given by {\tt SSE} code \citep{huretal2000}. This evolutionary sequence was incorporated within a numerical integration code that propagates planetary orbits. The details of this code are described in \cite{musetal2018}. Here, we used its RADAU integrator, and adopted a tolerance of $10^{-12}$. We started the simulations at the end of the main-sequence phase and ran the system for 10 Myr to clear out a minimum number of asteroids which would not have survived the main-sequence. The simulations then continued through the giant branch phases, and then for 1 Gyr during the white dwarf phase. In total, the simulation duration corresponds to 1.343 Gyr.

We henceforth use the word ``instability" to describe a simulation where the asteroid has escaped the system, collided with the planet, or collided with the star. These outcomes contain quantitative subtleties: the boundaries of the system here are assumed to be aspherical according to the Galactic tide, and are changing with mass loss from the evolving star \citep{vereva2013,veretal2014c}. These boundaries form a triaxial ellipsoid, one with semi-axes on the order of $10^5$ au, by assuming a Galactocentric distance of 8 kpc. Collisions with the central star were computed by assuming that after the star becomes a white dwarf, its Roche sphere is $1R_{\odot}$: within the code, we hence artificially inflated the star's radius to this typical value\footnote{Any object reaching the Roche radius will eventually be accreted onto the star itself, unless the object has an unusually high internal strength \citep{broetal2017,mcdver2021} or is sufficiently large such that its breakup produces a widely dispersed range of energies \citep{malper2020a}.}. 

\newpage

\section{Simulation results}

Although many aspects of the simulation outcomes may be analyzed, in this investigation we focus solely on the resonant-based outputs. A description of other properties and results, with similar but much lower resolution simulations, can be found in \cite{veretal2021}. We start here by presenting the full instability portraits (Section 3.1), and then zoom into three different locations: the regime of non-crossing orbits (Section 3.2), the $2$:$1$ resonance (Section 3.3) and the other first-order resonances (Section 3.4). Then, we compare libration widths computed with different sets of initial conditions in Section 3.5, and quantify different resonant strengths and instability outcomes in Section 3.6. 

\subsection{Full profiles}

\begin{figure*}
\includegraphics[width=15.0cm]{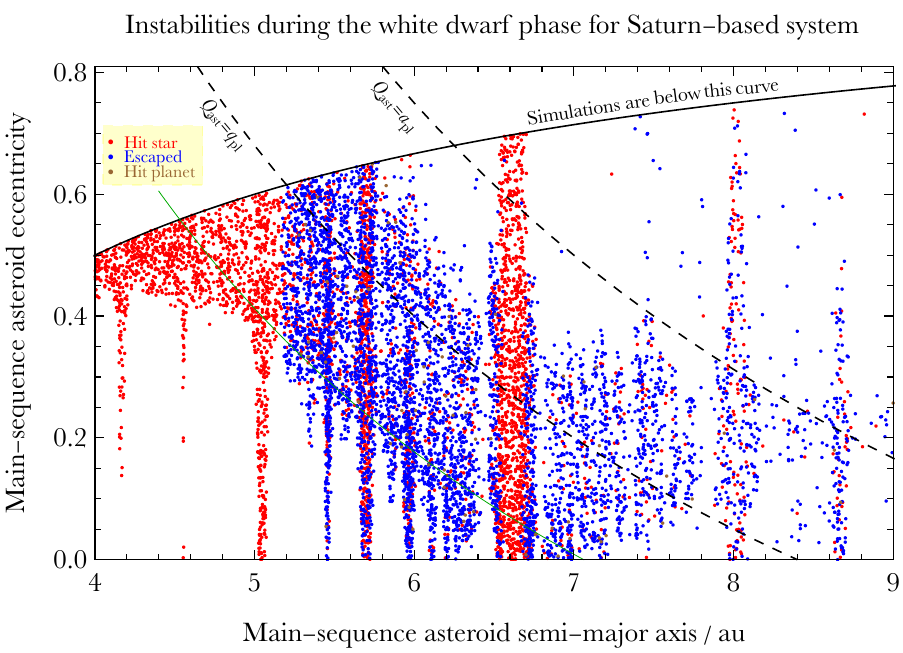}
\centerline{}
\includegraphics[width=15.0cm]{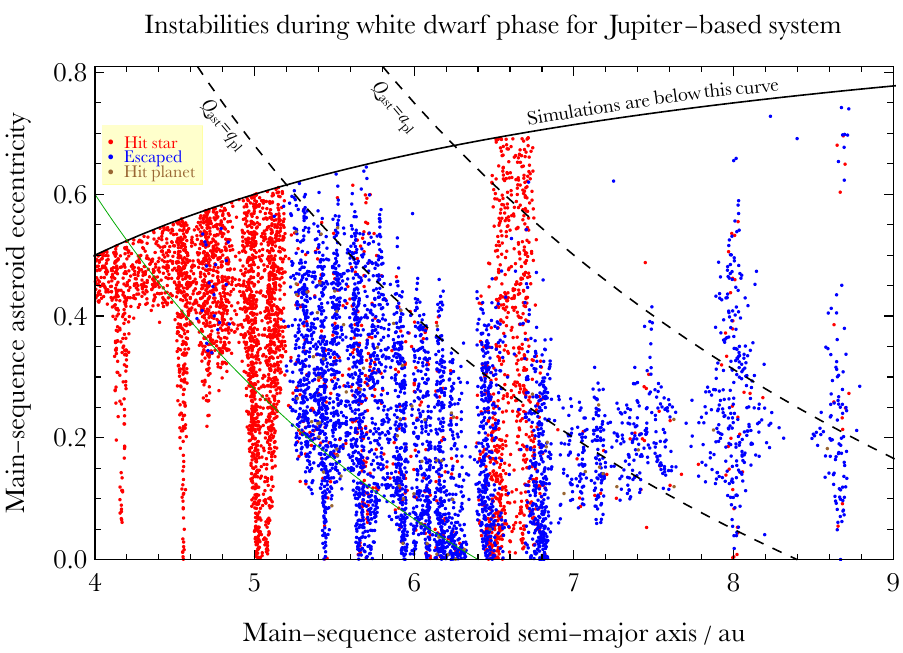}
\caption{
Full resonant instability portraits during the white dwarf phase {\rev only}, for the Saturn-based architecture (top) and the Jupiter-based architecture (bottom). {\rev The planet's main-sequence semi-major axis is 10.5 au and its eccentricity is 0.2. The black dashed curves indicate where the initial apocentre of the asteroid orbit intersects the initial semimajor axis and pericentre of the planet's orbit. The solid green curves (starting below the legends) from equation (\ref{HillEnc}) trace out purely analytic close encounter bounds between the planet and asteroids.}
}
\label{FullWD}
\end{figure*}

\begin{figure*}
\includegraphics[width=15.0cm]{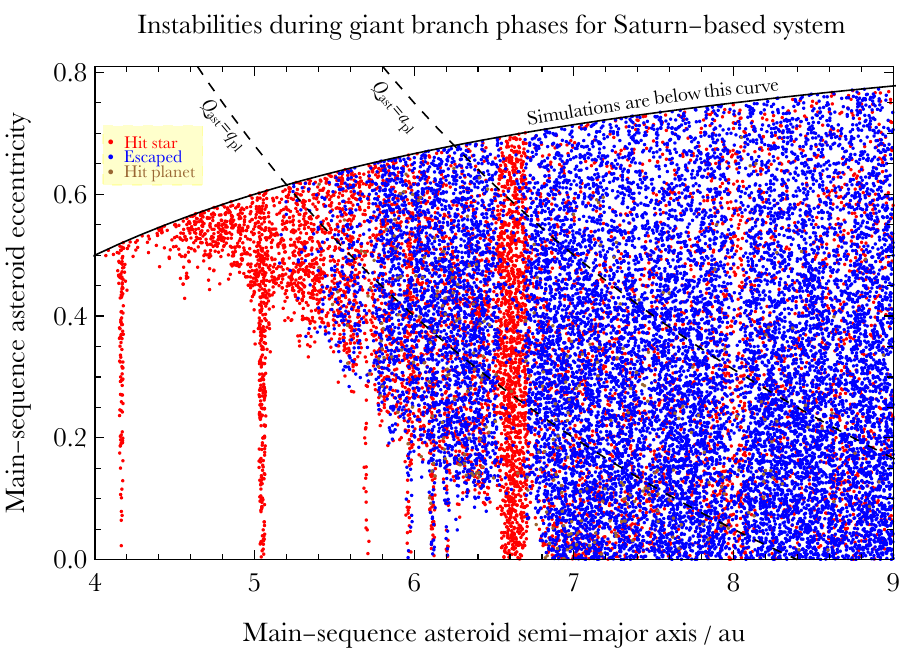}
\centerline{}
\includegraphics[width=15.0cm]{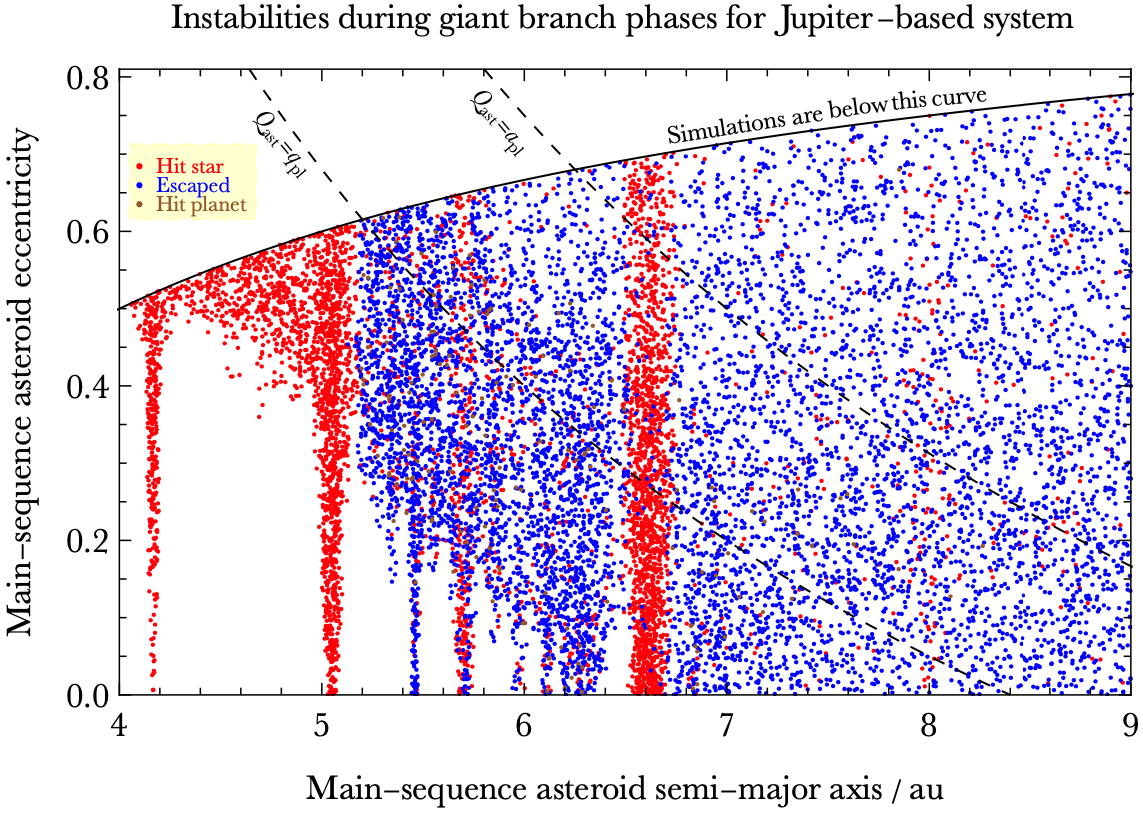}
\caption{
Full resonant instability portraits during the giant branch phases, for the Saturn-based architecture (top) and the Jupiter-based architecture (bottom). 
}
\label{FullGB}
\end{figure*}

\begin{figure*}
\includegraphics[width=15.0cm]{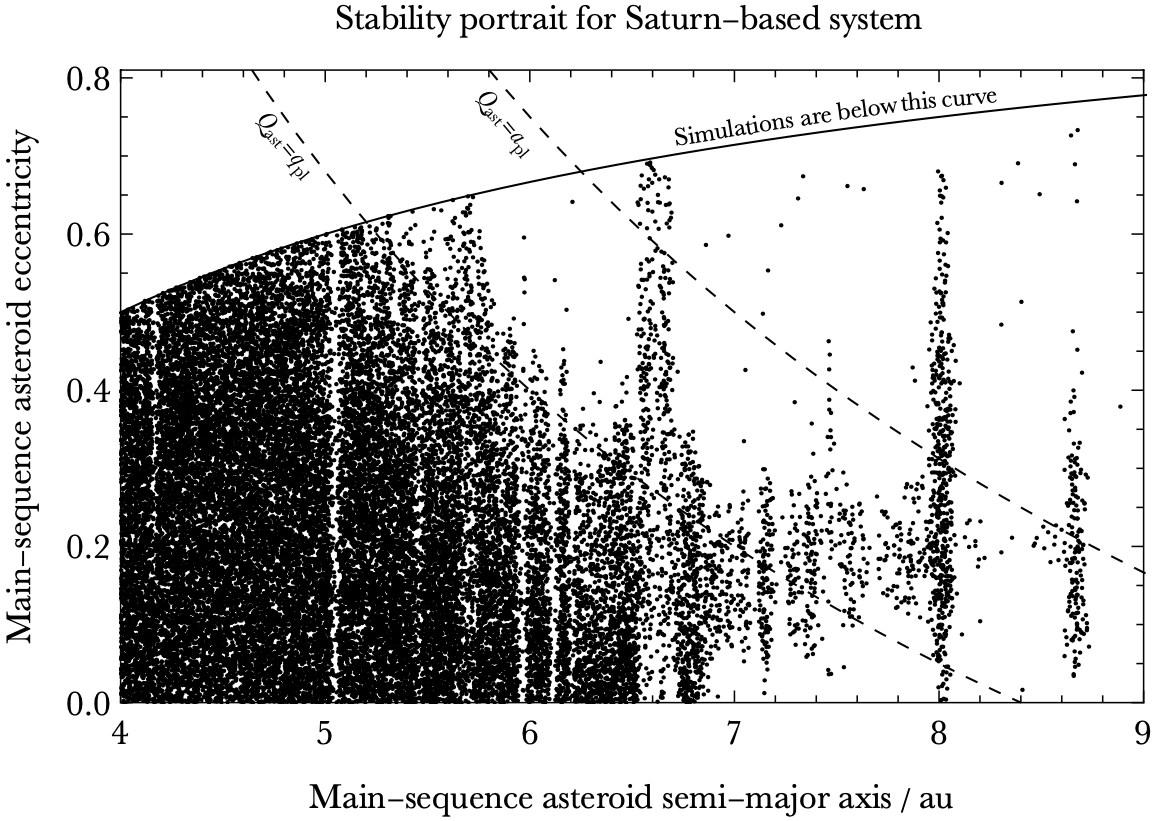}
\centerline{}
\includegraphics[width=15.0cm]{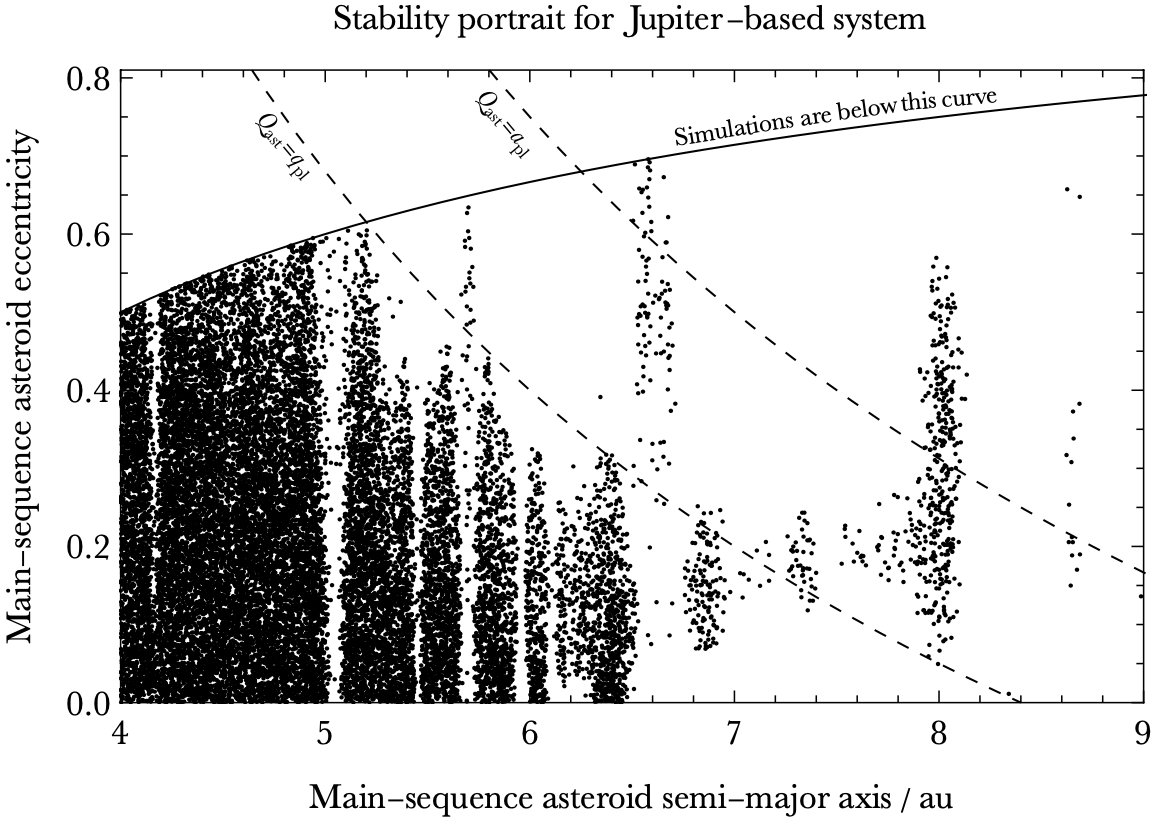}
\caption{
Full resonant stability portraits for the Saturn-based architecture (top) and the Jupiter-based architecture (bottom). All these asteroids remained stable {\rev throughout the numerical simulations until they ended,} after 1 Gyr of white dwarf cooling. {\rev The orbit-crossing curves appear to bound some of the stable regions.}
}
\label{FullStable}
\end{figure*}

We display the complete instability and stability profiles of our simulations in Figs. \ref{FullWD}-\ref{FullStable} without yet introducing any markings which are indicative of resonance. In each figure, the top panel is for the system with the Saturn-mass planet, and the bottom panel is for the system with the Jupiter-mass planet.

Figures \ref{FullWD} and \ref{FullGB} identify the asteroids which became unstable during the, respectively, white dwarf and giant branch phases, and Fig. \ref{FullStable} identifies the stable asteroids. In the instability plots, the three colours of dots represent different instability outcomes. We present these profiles, as well as most of the plots in this paper, as a function of both $e_{\rm ast}$ and $a_{\rm ast}$, which represent initial conditions (corresponding to a time which is 10 Myr before the end of the main-sequence).

Each plot contains $3-4$ curves: one solid {\rev black}, two dashed {\rev black,} {\rev and potentially one solid green}. The {\rev black} solid curve reflects the ($a_{\rm ast}, e_{\rm ast}$) boundary above which no asteroids were simulated. The dashed line labelled $Q_{\rm ast} = q_{\rm pl}$ represents the orbit crossing boundary when the arguments of pericentre of the orbits are anti-aligned: above this dashed line, the initial apocentre of the asteroid ($Q_{\rm ast}$) exceeds the {\rev initial} pericentre of the planet ($q_{\rm pl}$). The dashed line labelled $Q_{\rm ast} = a_{\rm pl}$ instead represents the location where the initial apocentre of the asteroid equals the {\rev initial} semi-major axis of the planet. 

{\rev The solid green curve represents an analytical approximation of the close encounter interactions between the planet and the asteroids during the white dwarf phase, where we assume that a close encounter occurs within three times the Hill radius of the pericentre of the planet's orbit. The curve is hence given by the following function:

\[
e_{\rm ast}(a_{\rm ast}) = \frac{q_{\rm pl}^{\rm WD} - 3R_{\rm Hill}^{\rm WD}}{a_{\rm ast}^{\rm WD}} - 1
\]
\[
\ \ \ \ \ \ \ \ \ \ \
            = \frac{\left(\frac{M_{\star}^{\rm MS}}{M_{\star}^{\rm WD}}\right) q_{\rm pl} - 
                         3  \left( 
                            \frac{M_{\star}^{\rm MS}}{M_{\star}^{\rm WD}}
                      \right)^{4/3} R_{\rm Hill}^{\rm MS}}
               {      \left( 
                            \frac{M_{\star}^{\rm MS}}{M_{\star}^{\rm WD}}
                      \right)
                      a_{\rm ast}
                      }
                   - 1
                   \]
\begin{equation}
\ \ \ \ \ \ \ \ \ \ \ = \frac{q_{\rm pl}}{a_{\rm ast}}
          \left[1 - 3^{2/3} \left( \frac{M_{\rm pl}}{M_{\star}^{\rm WD}} \right)^{1/3}  \right]
          - 1
          \label{HillEnc}
\end{equation}

\noindent{}where $R_{\rm Hill}^{\rm MS}$ and $R_{\rm Hill}^{\rm WD}$ are the Hill radii of the planet during the main-sequence and white dwarf phases respectively; the 4/3-power relation between these quantities was derived in \cite{payetal2016}. Interestingly, equation (\ref{HillEnc}) is independent of $M_{\star}^{\rm MS}$, which may be useful when modelling observed white dwarf planetary systems. All of the curves in Figs. \ref{FullWD}-\ref{FullStable} will also be drawn} in other plots in this paper.

Figures \ref{FullWD}-\ref{FullStable} have {\rev five} immediately apparent features. The first is that in almost no case do the asteroids collide with the planet: the instabilities are instead dominated by collisions with the star and escape from the system. The second is that asteroids in the non-resonant regions corresponding to $a_{\rm ast} = 6.6-9.0$ au are largely cleared out through escape during the giant branch phases. The third is that {\rev for the white dwarf phase, the green solid curves appear to usefully bound some of the unstable regions\footnote{{\rev Also, not shown is how the main-sequence analogue of equation (\ref{HillEnc}) accurately demarcates the boundary where asteroids are protected from the initial 10 Myr clear-out phase during main-sequence evolution.}}. The fourth is that} asteroids in the non-resonant regions corresponding to $a_{\rm ast} = 4.0-6.6$ au remain mostly stable. The {\rev fifth} is that resonances can act to either stabilize or destablize asteroids. We will quantify these trends in Section 3.6.

Comparing the two plots in each figure reveals other features. The Jupiter-mass planet clears away more asteroids than the Saturn-mass planet, as expected. The orbit-crossing dashed curve appears to play a role in carving out the {\rev long-term} stability boundaries, despite the initial values of $\omega_{\rm ast}$ being randomized. This dashed curve also serves as a useful demarcation for our subsequent analysis, which now begins with resonances below this curve.

\subsection{Resonances for non-crossing orbits}

\begin{figure*}
\includegraphics[width=14.0cm]{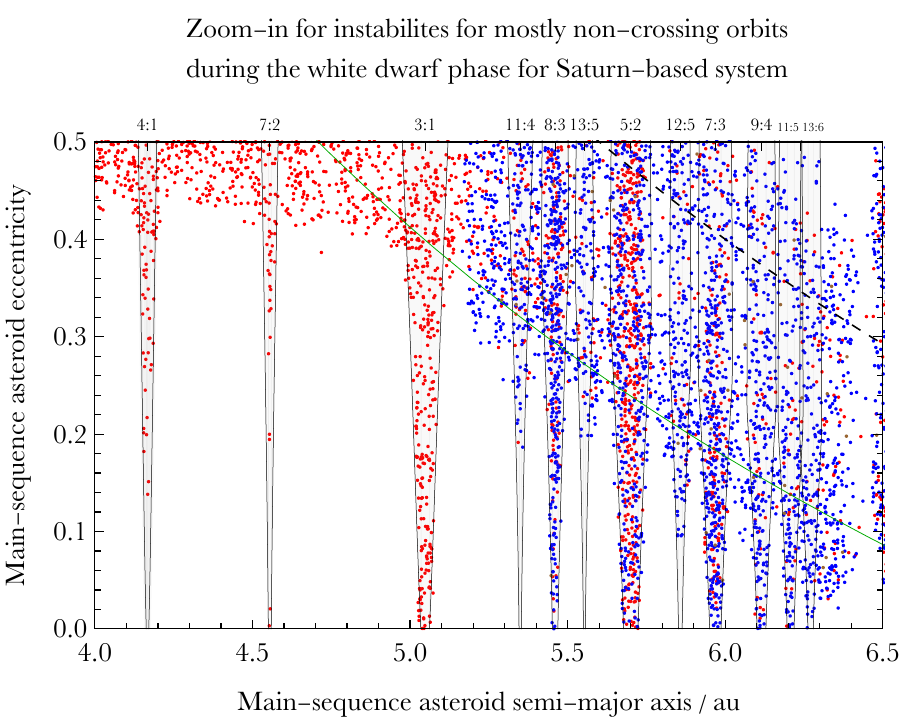}
\centerline{}
\includegraphics[width=14.0cm]{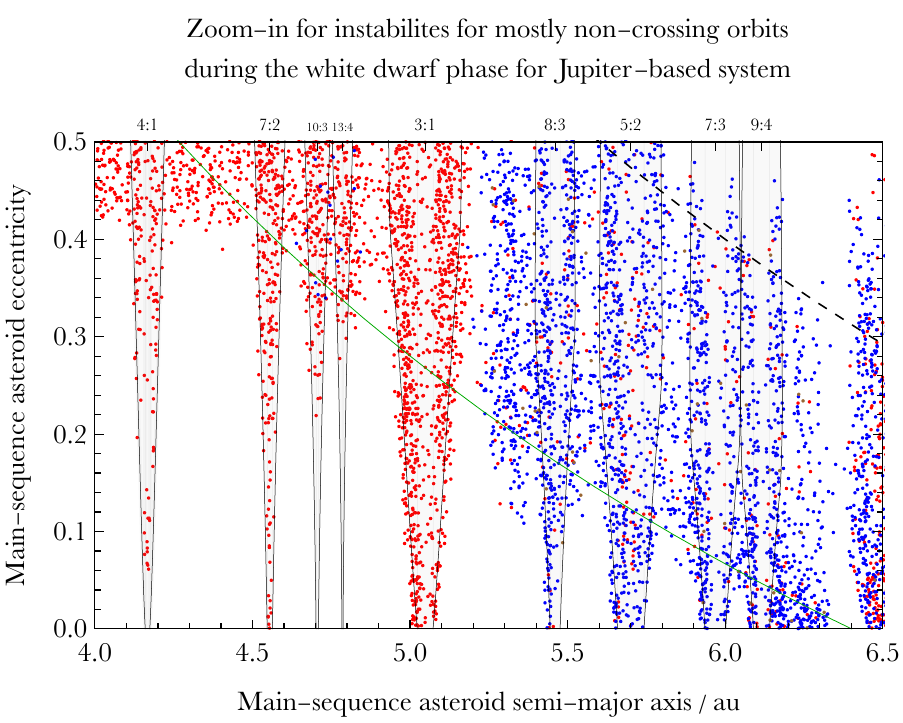}
\caption{
Resonant structures (in light gray shadings) for {\rev primarily} non-crossing orbits, for the Saturn-based architecture (top) and the Jupiter-based architecture (bottom). The maximum libration widths were computed independently of the $N$-body simulations by integrating with the semi-analytic numerical code of Gallardo et al. (2021). These maximum libration regions are roughly triangular in shape in the non-crossing orbit regime.
}
\label{Zoom1}
\end{figure*}
   
Here we focus on resonant structures that are contained within non-crossing orbit regions, which are highlighted in Fig. \ref{Zoom1}.

The plots in that figure represent zoom-ins of the plots in Fig. \ref{FullWD}, with superimposed libration regions represented by the light gray shaded areas and bound by pairs of thin black curves. These libration widths are computed from \cite{galetal2021} and actually represent {\it maximum} libration widths. 

In order to compute these maximum widths, for the star, we assumed a {\rev stellar} mass of $2.0M_{\odot}$. For the planet, we adopted $a_{\rm pl} = 10.5$ au, $e_{\rm pl} = 0.2$, and $i_{\rm pl} = \varpi_{\rm pl} = \Omega_{\rm pl} = 0^{\circ}$, where $\varpi_{\rm pl}$ represents the longitude of pericentre of the planet. For the asteroid, we chose $a_{\rm pl}$ according to the location of the nominal resonance ($=10.5 \, {\rm au} \left(p/q\right)^{-2/3}$, for a $p$:$q$ resonance), and sampled $e_{\rm ast}$ in increments of 0.025. We then fixed $i_{\rm ast} = \Omega_{\rm ast} = 0^{\circ}$, but for each value of $e_{\rm ast}$, sampled 180 equally-incremented values of $\varpi_{\rm ast}$ starting at $0^{\circ}$. We finally adopted the maximum width attained from these 180 values, and connected these maximum values at different eccentricities with interpolated curves. 

Therefore, the computation of maximum libration widths did not explicitly take into account stellar evolution or the evolution of the asteroid's orbital parameters. Nevertheless, this method of computing maximum libration widths with just main-sequence values appears to be effective when compared to the results of the $N$-body simulations -- which are completely independent from the code used to compute the libration widths. 

{\rev As can be seen, these widths accurately reflect the resonant influences that are indicated by the outcome of the $N$-body simulations. These widths all} monotonically increase with $e_{\rm ast}$ until around the eccentricity value where the orbits cross (the dashed line). Around this location, the widths {\rev become nearly fixed or slightly decrease for greater $e_{\rm ast}$}. How reflective this structure is of the actual physical evolution is difficult to determine because at {\rev high} values of $e_{\rm ast}$, resonant overlap occurs more readily and the unstable asteroids are sparse and do not form a coherent structure.

{\rev For crossing orbits, the libration widths are dependent on the close encounter criterion set in the \cite{galetal2021} code through the parameter {\tt rhtol}. \cite{gallardo2020} found that an accurate value for this close encounter criterion in the restricted three-body problem is three planetary Hill radii ({\tt rhtol} $=3$), which we adopted here. We also performed extensive tests with {\tt rhtol} $=0$ and confirmed that while the libration widths with that value do not differ much for $e_{\rm ast}$ below the collision curve, above the collision curve widths become unrealistic and do not reflect the outcome of our $N$-body simulations.}

Comparing the two plots in Fig. \ref{Zoom1} reveals that during the white dwarf phase, the influence of some resonances vanish when the planetary mass changes; only resonances which have a discernible effect on the dynamics are plotted. For example, during this phase of stellar evolution, the Jupiter-mass planet apparently accessed pollution reservoirs at the 7th- and 9th-order resonances $10$:$3$ and $13$:$4$; such reservoirs are not easily identifiable for the Saturn-mass planet architecture.

\subsection{The $2$:$1$ resonance}   

\begin{figure*}
\includegraphics[width=14.0cm]{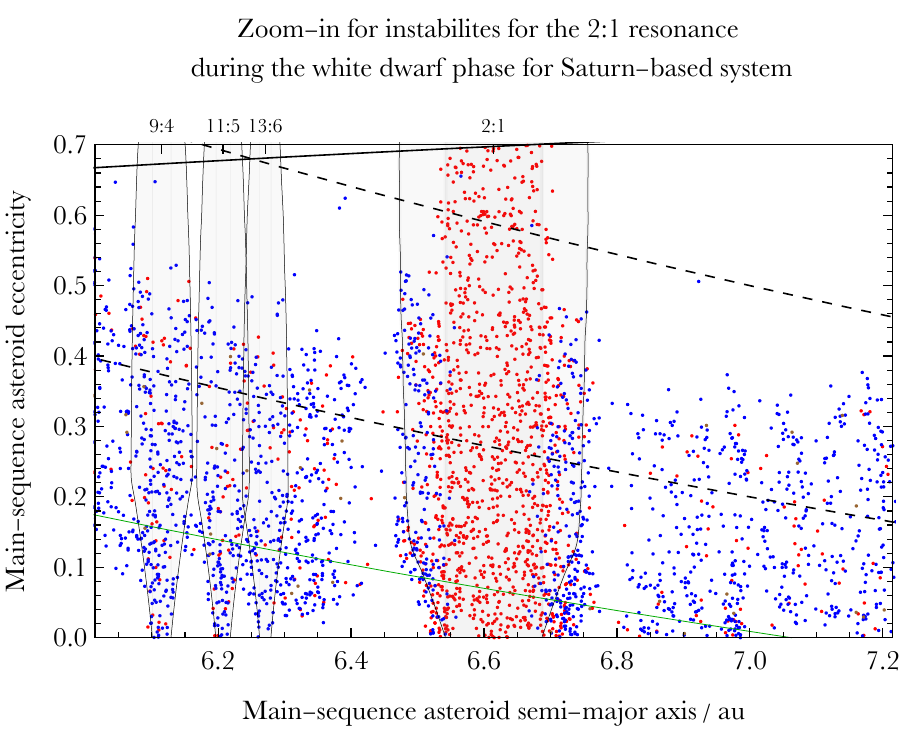}
\centerline{}
\includegraphics[width=14.0cm]{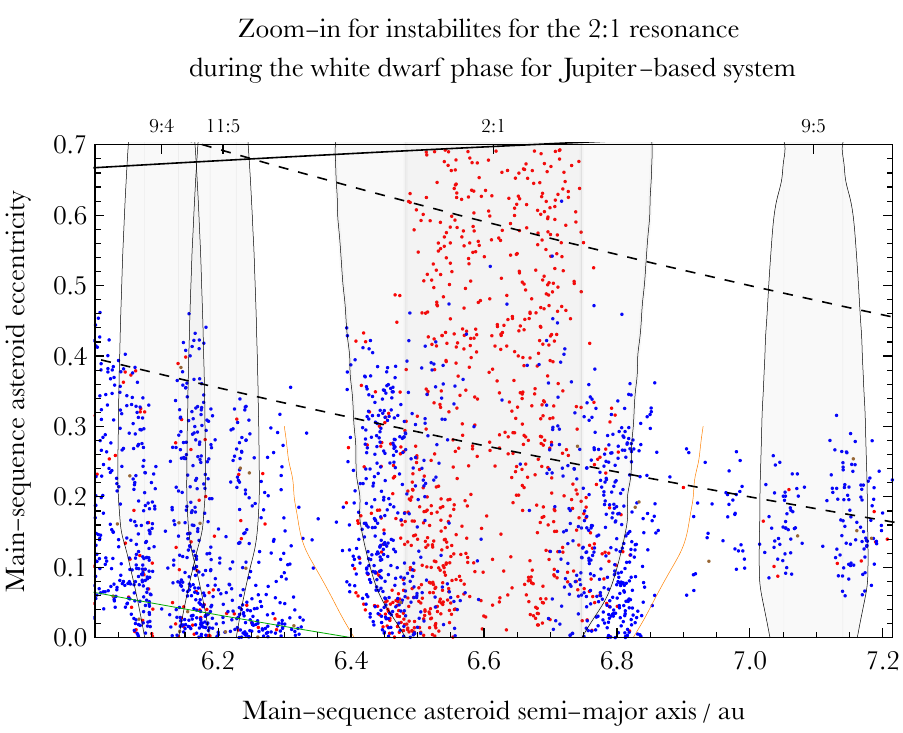}
\caption{
The $2$:$1$ resonant structures for the Saturn-based (top) and Jupiter-based (bottom) architectures. In the orbit-crossing regime (above the lower dashed curve) the semianalytic maximum libration structure {\rev stops increasing monotonically with $e_{\rm ast}$}. The $2$:$1$ resonant core (the slightly darker gray regions) usefully {\rev separate white dwarf pollutants from escapers} at high $e_{\rm ast}$.
}
\label{Zoom2}
\end{figure*}

We next zoom into the region surrounding the $2$:$1$ resonance, in Fig. \ref{Zoom2}. This resonance deserves special attention both because of its strength -- being a first-order resonance -- and because it has been highlighted in previous post-main-sequence resonance studies. {\rev Also, Fig. \ref{FullWD} indicates immediately that the $2$:$1$ resonance is a significant driver of white dwarf pollution.}

In Fig. \ref{Zoom2}, there are resonant structures which are immediately apparent aside from the $2$:$1$ resonance. Libration widths for these other resonances have been computed and displayed on the plots, although their influence is clear {\rev primarily} in the non-crossing orbit regime. For the $2$:$1$ resonance, for the planetary architecture that we adopted, crossing orbits occur when $e_{\rm ast} \gtrsim 0.27$. 

We see structure on both sides of this eccentricity boundary. For $e_{\rm ast} \lesssim 0.27$, the computed maximum libration width region {\rev does} provide a qualitatively-good bound on resonant behaviour, but does {\rev appear to slightly} underestimate the true region of influence. 

One potential reason for the underestimation is that the resonance width computed here does not take into account stellar evolution. When the star-to-planet mass ratio changes, the libration width should change also. \cite{debetal2012} explicitly drew this change for the $2$:$1$ resonance by using the formula for libration width from \cite{murder1999}, which does not incorporate as many input parameters as the semianalytical prescription from \cite{galetal2021}.

Hence, as an exploratory exercise, for the Jupiter-based system, we recomputed the maximum libration width with the \cite{galetal2021} prescription, but instead adopted the white dwarf mass of $0.637M_{\odot}$ instead of the main-sequence mass of $2.0M_{\odot}$. We drew the result as {\rev a pair of} thin superimposed {\rev solid orange} curves on the bottom plot of Fig. \ref{Zoom2}. These curves, however, appear to significantly overestimate the region of influence of the resonance {\rev for $e_{\rm ast} \gtrsim 0.05$.}

Returning to our original computation of libration width, the instability types suggest that we may usefully split the resonant region {\rev above} the orbit-crossing curve into a ``core" and a ``tail". We define the core as the width computed at $e_{\rm ast} = 0$ as applied to all $e_{\rm ast}$, as indicated by the {\rev slightly darker gray region}. The tail is then defined as the region outside of the core but bounded by libration width boundaries.

With these definitions, a visual inspection of the plot immediately indicates that $2$:$1$ asteroids which pollute the white dwarf are located in the resonant core, and those which escape the system are located in the tails. This trend appears to continue even in the crossing-orbit regime ($e_{\rm ast} \ge 0.27$) except that the locations of the asteroids which feature escape stick to the boundaries of the core before vanishing entirely above some critical eccentricity. These trends hold for both the Jupiter-mass and Saturn-mass cases. 

{\rev The physical explanation for this behaviour is that only those asteroids which are deep enough in the resonance (in the core) can secularly evolve such that their eccentricities are increased to near-unity while their semi-major axes remain nearly constant. Alternatively, asteroids that are on the edge of the resonance (in the tails) more easily escape the resonance and subsequently experience encounters with the planet, allowing their semi-major axes to be periodically pumped until ejection.}

\subsection{The $3$:$2$ and $4$:$3$ resonances}   
   
\begin{figure}
\includegraphics[width=8.5cm]{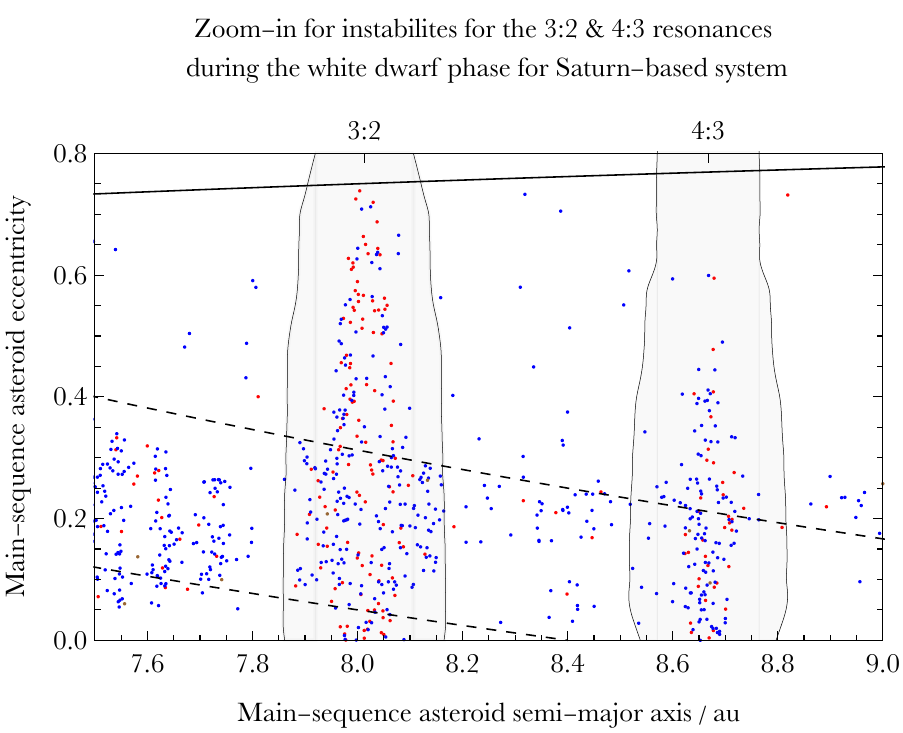}
\centerline{}
\includegraphics[width=8.5cm]{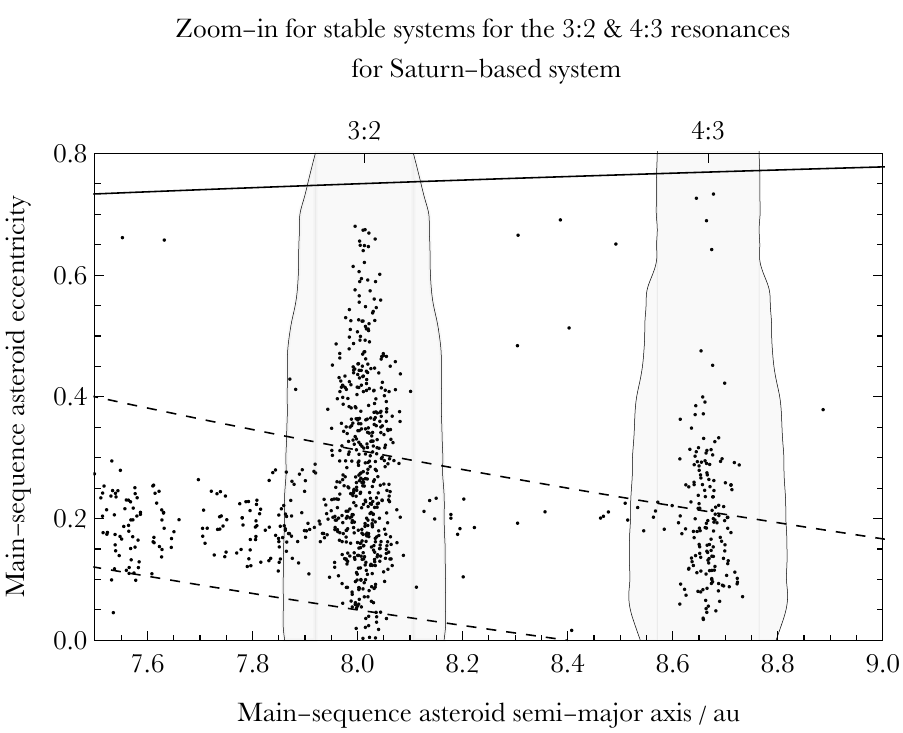}
\caption{
Influence of the $3$:$2$ and $4$:$3$ resonances, which are located almost entirely in the crossing-orbit regime, on instability location during the white dwarf phase (top panel) and on maintaining stability (bottom panel). Both resonances clearly influence the stability portraits across the sampled parameter space.
}
\label{Zoom3}
\end{figure}

{\rev Next}, we zoom-in to the rightmost portion of Figs. \ref{FullWD} and \ref{FullStable} in Fig. \ref{Zoom3}, which features asteroids that are almost exclusively in the crossing-orbit regime. Here, {\rev for the two first-order resonances,} the maximum libration width computed from the \cite{galetal2021} prescription {\rev does not increase monotonically with $e_{\rm ast}$, but rather does just the opposite.}

The distribution of both unstable and stable asteroids which are illustrated in the figure strongly suggest that both {\rev the $3$:$2$ and $4$:$3$} resonances have played a role in the asteroids' dynamical evolution. The $3$:$2$ resonance appears to have exerted influence on asteroids with higher values of $e_{\rm ast}$ than the $4$:$3$ resonance. Both resonance {\rev regions} include a mixture of asteroids which pollute the white dwarf, and those which escape from the system during the white dwarf phase.

The bottom panel of Fig. \ref{Zoom3} is notable for how it demonstrates that asteroids can be protected from instability through different stages of stellar evolution by being trapped into a strong resonance. The plot also shows how such asteroids are more likely to be protected for low values of $e_{\rm ast}$, even though a few remain stable at nearly the highest values of $e_{\rm ast}$ sampled.

\subsection{Libration width comparisons}

We have now shown, by comparison with $N$-body simulations, that the maximum libration regions computed with main-sequence values may be sufficiently accurate to be used as reliable proxies for the initial location of post-main-sequence resonant dynamical behaviour. In this subsection, we explore in more detail how these maximum libration widths vary depending on the input parameters which are adopted. {\rev All plots in this section show $N$-body simulation output from the white dwarf phase only.}

\begin{figure}
\includegraphics[width=8.5cm]{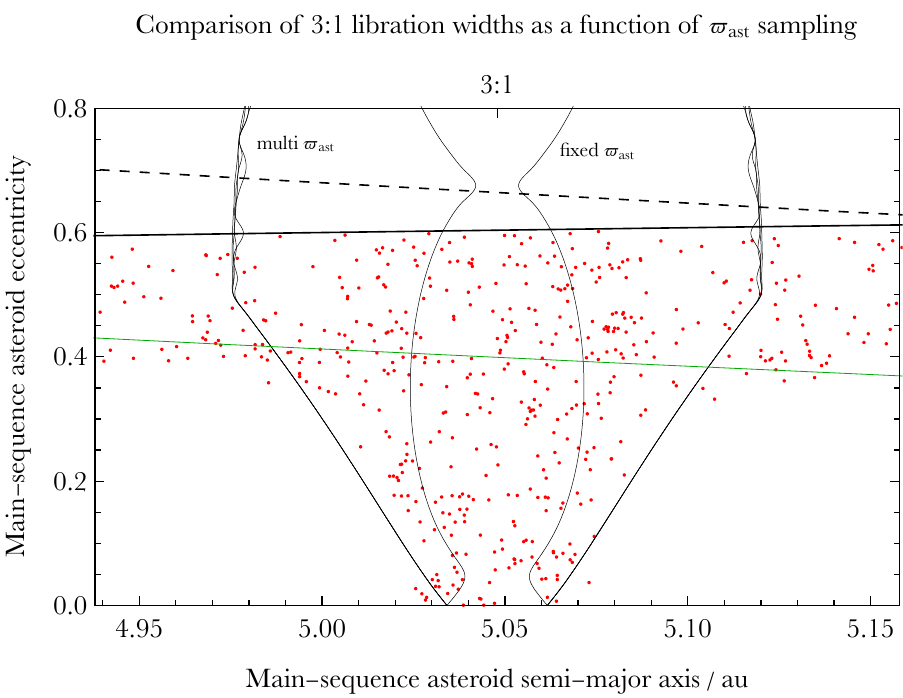}
\caption{
Result of computing maximum libration widths for the $3$:$1$ resonance with five different sampling frequencies for $\varpi_{\rm ast}$ (with 180, 72, 36, 18, 1 values sampled, starting at $0^{\circ}$). This plot illustrates that the maximum libration width is largely independent of sampling frequency. However, adopting a fixed value of $\varpi_{\rm ast}$ generates a qualitatively differently-shaped region (middle pair of curves). 
}
\label{Zoom31WDvarpi}
\end{figure}

First, we tested the sensitivity of our maximum libration width computations on the sampling frequency of $\varpi_{\rm ast}$. So far, we have computed widths for each of 180 equally-spaced values of $\varpi_{\rm ast}$, and then drew output corresponding to the maximum of these widths. Figure \ref{Zoom31WDvarpi} illustrates the result of reducing this sampling rate for the $3$:$1$ resonance in the Saturn-based system.

On the figure, we drew five different pairs of curves corresponding to sampling rates of 180, 72, 36, 18 and 1 value. The differences between the first four pairs of curves are not discernible {\rev for most of} the non-orbit crossing regime, helping to illustrate the robustness of our fiducial choice for the plots in this paper. However, when we sampled just a single value of $\varpi_{\rm ast}$, in this case $\varpi_{\rm ast} = 0^{\circ}$, the result differs, as shown by the middle pair of curves.

Overall then, this example demonstrates that adopting a single value of $\varpi_{\rm ast}$ produces a quantitatively different, and more inaccurate, result than sampling multiple values and taking the maximum output. Further, the results of our $N$-body simulations demonstrate that the libration region in the non-orbit crossing regime increases monotonically with $e_{\rm ast}$, and does not curve inward {\rev at $e_{\rm ast}=0.05$}, as indicated by the $\varpi_{\rm ast} = 0^{\circ}$ curve.

\begin{figure}
\includegraphics[width=8.5cm]{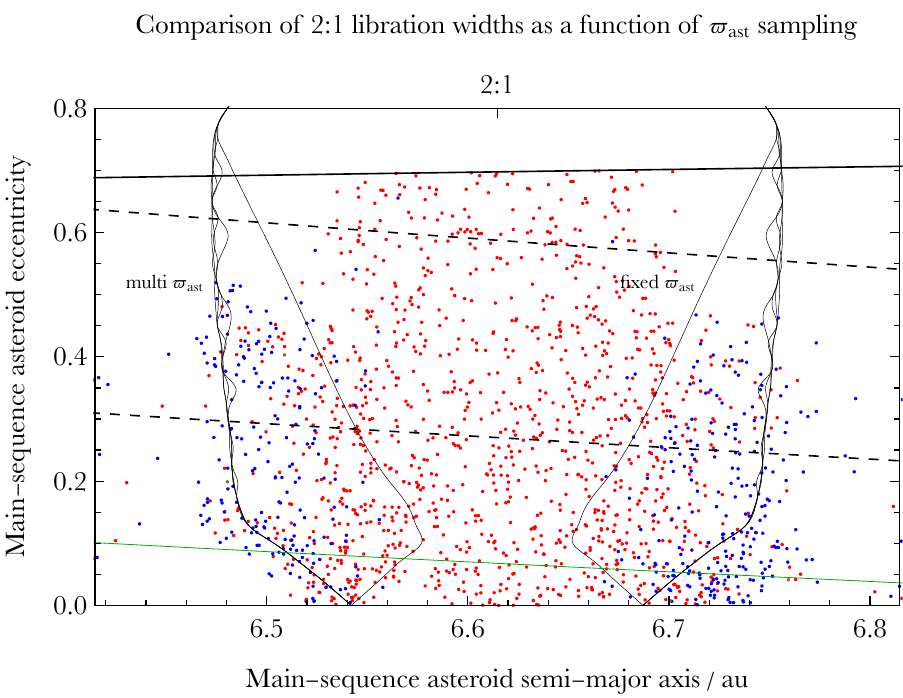}
\caption{
Similar to Fig. \ref{Zoom31WDvarpi}, except for the $2$:$1$ resonance, where the result of adopting a fixed value of $\varpi_{\rm ast}$ is more pronounced and would lead to a more significant underestimation of the number of asteroids which are influenced by this resonance.
}
\label{Zoom21WDvarpi}
\end{figure}

In order to fortify these conclusions, we performed the same exercise for the $2$:$1$ resonance. The result is shown in Fig. \ref{Zoom21WDvarpi}, and is the same: the shape of the curves resulting from the fixed value of $\varpi_{\rm ast} = 0^{\circ}$ does not reflect the region of influence from the $N$-body simulations, and here significantly underestimates the number of asteroids under this influence.

\begin{figure}
\includegraphics[width=8.5cm]{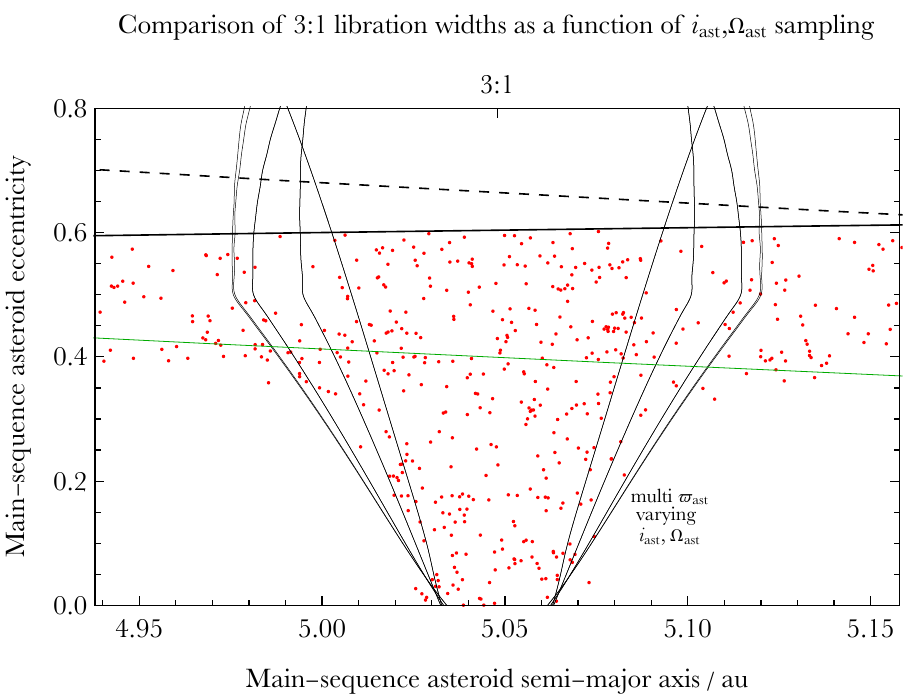}
\caption{
Result of computing maximum libration widths for the $3$:$1$ resonance with different initial values of ($i_{\rm ast}$, $\Omega_{\rm ast}$). The different pairs adopted, in addition to the nominal pair ($0^{\circ}$, $0^{\circ}$), were ($5^{\circ}$, $0^{\circ}$), ($20^{\circ}$, $0^{\circ}$), ($20^{\circ}$, $180^{\circ}$), ($50^{\circ}$, $0^{\circ}$), ($50^{\circ}$, $90^{\circ}$), ($50^{\circ}$, $180^{\circ}$) and ($50^{\circ}$, $270^{\circ}$). 
}
\label{Zoom31WDiOmega}
\end{figure}

Because the libration width prescription of \cite{galetal2021} requires a full set of three-dimension orbital inputs for both the asteroid and planet, and because all orbital elements of the asteroids varied throughout our simulations, we then explored how these libration curves change by varying $i_{\rm ast}$ and $\Omega_{\rm ast}$. The result is shown in Fig. \ref{Zoom31WDiOmega}.

Illustrated in this figure are 8 different pairs of maximum libration width curves, some of which overlap, that correspond to ($i_{\rm ast}$, $\Omega_{\rm ast}$) = ($0^{\circ}$, $0^{\circ}$), ($5^{\circ}$, $0^{\circ}$), ($20^{\circ}$, $0^{\circ}$), ($20^{\circ}$, $180^{\circ}$), ($50^{\circ}$, $0^{\circ}$), ($50^{\circ}$, $90^{\circ}$), ($50^{\circ}$, $180^{\circ}$) and ($50^{\circ}$, $270^{\circ}$). All of these curves do present shapes which are reflective of the results of the $N$-body simulations, but with slightly different tails. 

For asteroids that pollute the white dwarf, we expect that the extent of the isotropy with which asteroids enter the white dwarf's Roche radius to increase with planet mass \citep{veretal2021}. Hence, for particularly massive planets, we might obtain more accurate bounds on the libration region by sampling multiple values of $i_{\rm ast}$, $\Omega_{\rm ast}$ and $\varpi_{\rm ast}$, rather than just $\varpi_{\rm ast}$. However, doing so would become computationally expensive.

\begin{table*}
 \centering
 \begin{minipage}{180mm}
   \centering
   \caption{Simulation outcomes for selected resonant regions; WD stands for white dwarf and GB stands for giant branch. Included are all 1st- to 5th-order resonances between the $4$:$1$ and $2$:$1$ inclusive, as well as the $3$:$2$ and $4$:$3$ resonances, which are both deep into the crossing orbit regime. The value of $a_{\rm ast, nom}$ represents the asteroid's nominal initial location for each resonance. The underlined values in the {\rev fourth} column highlight relatively efficient white dwarf polluters (the $2$:$1$, $3$:$1$ and $4$:$1$ resonances). 
}
   \label{OutcomeTable}
  \begin{tabular}{lccccccccc}
\hline
 Resonance  & $a_{\rm ast,nom}$ & {\rm Total} & {\rm WD} & {\rm WD} & {\rm WD} & {\rm GB} & {\rm GB} & {\rm GB} & {\rm Stable} \\[2pt]
   & / au &  {\rm Number} & {\rm Hit star} & {\rm Escaped} & {\rm Hit planet} & {\rm Hit star} & {\rm Escaped} & {\rm Hit Planet} &  \\[2pt]
\hline
\hline

$4$:$1$ Saturn  & 4.17  & 531  & $\underline{21\%}$  &  $0\%$  &  $0\%$    & $14\%$  & $0\%$   & $0\%$     &  $55\%$  \\[2pt]
$4$:$1$ Jupiter & 4.17  & 1014  & $\underline{32\%}$  &  $0\%$  &  $0\%$    & $16\%$  & $0\%$   & $0\%$     &  $41\%$  \\[2pt]
$7$:$2$ Saturn  & 4.55  & 543  & $5.3\%$ &  $0\%$  &  $0\%$    & $13\%$  & $0\%$   & $0\%$     &  $81\%$  \\[2pt]
$7$:$2$ Jupiter & 4.55  & 900  & $7.1\%$ &  $0\%$  &  $0\%$    & $31\%$  & $0\%$   & $0\%$     &  $60\%$  \\[2pt]
$3$:$1$ Saturn  & 5.05  & 1666 & $\underline{24\%}$ &  $0\%$  &  $0\%$    & $24\%$  & $0\%$   & $0\%$     &  $40\%$  \\[2pt]
$3$:$1$ Jupiter & 5.05  & 2986 & $\underline{36\%}$  & $0.033\%$& $0\%$   & $33\%$  & $0.033\%$   & $0\%$     &  $16\%$  \\[2pt]
$8$:$3$ Saturn  & 5.46  & 1187 & $8.6\%$ & $2.8\%$ & $0.08\%$  & $14\%$  & $33\%$  & $0.25\%$  &  $32\%$  \\[2pt]
$8$:$3$ Jupiter & 5.46  & 1886 & $5.9\%$ & $22\%$  & $0.80\%$   & $2.2\%$ & $19\%$  & $0.42\%$  &  $14\%$  \\[2pt]
$5$:$2$ Saturn  & 5.70  & 2102 & $8.2\%$ & $4.0\%$ & $0.29\%$ & $25\%$  & $22\%$  & $0.24\%$ &  $31\%$  \\[2pt]
$5$:$2$ Jupiter & 5.70  & 3234 & $10\%$  &  $14\%$ &  $0.56\%$ & $3.5\%$ & $17\%$  & $0.093\%$  &  $14\%$  \\[2pt]
$7$:$3$ Saturn  & 5.97  & 1803 & $13\% $ & $14\%$ &  $0.50\%$ & $8.4\%$  & $22\%$  & $0.44\%$  &  $16\%$  \\[2pt]
$7$:$3$ Jupiter & 5.97  & 2642 & $2.2\%$ & $12\%$  & $0.26\%$  & $1.3\%$ & $11\%$  & $0.26\%$  &  $9.0\%$  \\[2pt]
$9$:$4$ Saturn  & 6.11  & 1506 & $13\%$  & $21\%$  &  $0.86\%$  & $3.1\%$ & $16\%$  & $0.13\%$  &  $12\%$  \\[2pt]
$9$:$4$ Jupiter & 6.11  & 2328 & $2.2\%$ & $14\%$  & $0.34\%$  & $1.5\%$ & $11\%$  & $0.21\%$  &  $5.4\%$  \\[2pt]
$2$:$1$ Saturn  & 6.61  & 5295 & $\underline{27\%}$ & $7.3\%$ & $0.23\%$  & $21\%$  & $4.9\%$ & $0.019\%$ &  $12\%$  \\[2pt]
$2$:$1$ Jupiter & 6.61  & 8787 & $\underline{18\%}$  & $4.3\%$ & $0.057\%$ & $8.2\%$  & $5.6\%$  & $0.023\%$  &  $3.4\%$  \\[2pt]
$3$:$2$ Saturn  & 8.01  &  6084 & $7.1\%$  & $18\%$  & $0.23\%$  & $1.9\%$   & $3.4\%$  & $0.033\%$  & $9.1\%$  \\[2pt]
$3$:$2$ Jupiter & 8.01  & 9619 & $1.3\%$ & $6.8\%$ & $0.031\%$ & $0.32\%$  & $2.5\%$ & $0.010\%$ & $4.0\%$  \\[2pt]
$4$:$3$ Saturn  & 8.67  & 5873 & $5.9\%$  & $23\%$  & $0.10\%$  & $0.63\%$  & $2.1\%$  & $0.034\%$  & $2.9\%$  \\[2pt]
$4$:$3$ Jupiter & 8.67  & 8881 & $0.53\%$ & $7.1\%$ & $0\%$ & $0.14\%$ & $1.1\%$ & $0.011\%$ &$0.17\%$  \\[2pt]

\hline
\end{tabular}
\end{minipage}
\end{table*}

\subsection{Resonant outcomes and strengths}   
   
So far, the plots in this paper have illustrated different instability outcomes for different resonances. In this subsection, we tabulate these results {\rev for selected resonances in Table \ref{OutcomeTable}.}

The resonances (1st column) {\rev are ordered} by horizontal location on Figs. \ref{FullWD}-\ref{FullStable} (second column). The {\rev third} column reports the total number of asteroids initially contained within that {\rev resonant} region. This number also includes those asteroids that have been cleared out during the first 10 Myr with a fixed stellar mass in order to allow the system to dynamically settle, at least to a minor extent.

The remaining columns report the outcomes of the simulations in percentages of the initial asteroid sample ({\rev third} column), with highlighted results underlined in the {\rev fourth} column. The table illustrates the following trends:

\begin{enumerate}

\item  The most significant polluters of white dwarfs arise from the $2$:$1$, $3$:$1$ and $4$:$1$ resonances, with at least {\rev 18} per cent of the asteroids in each of those resonant regions entering the Roche radius of the star. By way of comparison with another third-order resonance in the non-orbit crossing regime -- the $5$:$2$ -- the $4$:$1$ is a much more efficient polluter. This result supports the finding that the $4$:$1$ resonance can easily excite test particles to high eccentricity in the presence of a slightly eccentric planet \citep{picetal2017}.

{\rev
\item  A different measure of the efficiency of the $2$:$1$, $3$:$1$ and $4$:$1$ resonances to pollute the white dwarf is by comparison to the total number of polluters across our entire initial disc of asteroids. Although such a comparison is highly dependent on our choice of the inner and outer edges of the initial disc, this comparison may still be useful. For the Saturn-based and Jupiter-based systems, respectively, approximately $47$ and $80$ per cent of all polluters were generated from the $2$:$1$, $3$:$1$ and $4$:$1$ resonances. More specifically, for the Saturn-based system, each of these resonances provided respectively $35$, $9.6$ and $2.7$ per cent of pollutants, whereas for the Jupiter-based system, the percentages were $42$, $29$ and $8.6$.
}

\item  Resonances can stabilise asteroids, even those in crossing orbits (the $4$:$3$ and $3$:$2$), and sometimes in large quantities (tens of per cent of the original reservoir). These stable populations are also potentially important for white dwarf pollution, particularly at late times (beyond 1 Gyr of white dwarf cooling) {\rev because they} may be (a) accessed later by external perturbers which initially played no role in the dynamics within tens of au \citep{bonver2015,veretal2016}, or, (b) when close enough to the white dwarf, dragged into the star from its own radiation \citep{rafikov2011,veretal2015b,veretal2022,veras2020,malamudetal2021,broetal2022}. 

\item  Escape from the system is more difficult, if not impossible, from resonant regions close to the star. {\rev In this region close asteroid encounters with the planet are non-existent; such encounters would be required for an asteroid to be ejected.} 

\item  Asteroid collisions with the planet are rare in all cases. The three resonances which produce the greatest fraction of these types of collisions are the 4th- and 5th-order resonances $9$:$4$, $8$:$3$ and $7$:$3$, but only at the 1 per cent level.

\end{enumerate}

\begin{figure}
\includegraphics[width=8.5cm]{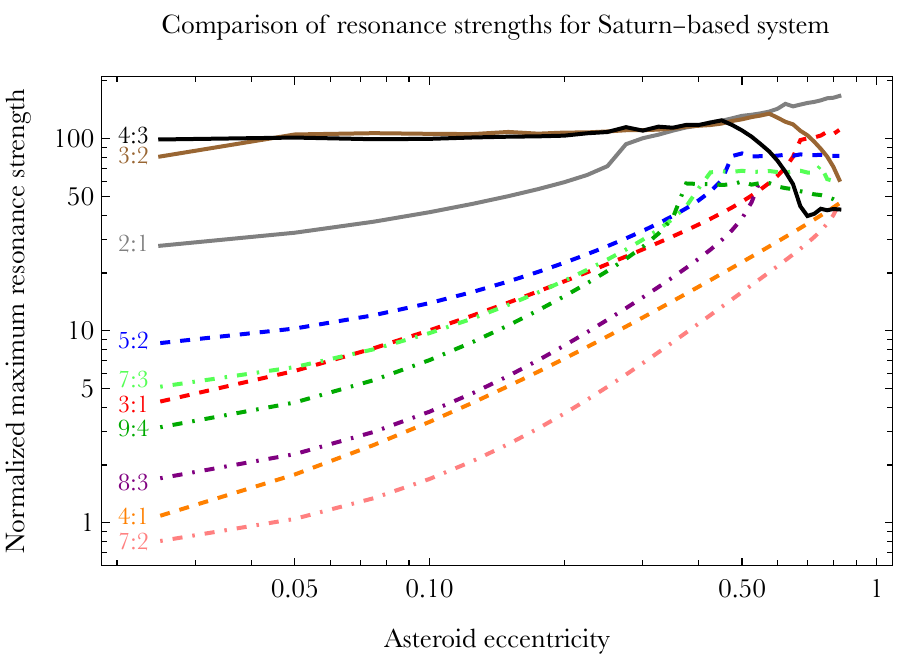}
\caption{
Relative strengths of different resonances as a function of $e_{\rm ast}$ for the Saturn-based system. The plot for the Jupiter-based system is equivalent, except for strengths which are scaled up by a factor of 3.34. The strengths are computed according to the prescription in Gallardo (2006). On this plot, the 1st-order resonances are given by solid curves, 2nd-order and 3rd-order resonances by dashed curves, and 4th-order and 5th-order resonances by dot-dashed curves. The curves are smooth {\rev and monotonically increasing with $e_{\rm ast}$} in the non-crossing orbit regime. 
}
\label{SaturnResStrength}
\end{figure}

In an attempt to quantify the strength of various resonances, we used a numerical strength measure from the semianalytic prescription that was defined in equation 8 of \cite{gallardo2006} as a difference in disturbing function evaluations, computed numerically (not with expansions). We then normalized this numerical strength measure for selected resonances as a function of $e_{\rm ast}$, and show the results in Fig. \ref{SaturnResStrength}.

That plot illustrates that in the non-crossing orbit regime, strength varies smoothly with $e_{\rm ast}$. Also, the differences in strength are most pronounced for asteroids on {\rev low-eccentricity} orbits. This strength metric usefully takes into account both resonant order and proximity to the planet, but {\rev does not trace secular evolution, the results of which are presented in} Table \ref{OutcomeTable}.

\section{Discussion}

The results of this investigation not only have intrinsic dynamical interest, but also may be applied to individual real systems in the near future, especially as more suitable one-planet polluted white dwarf systems are discovered. Right now no systems are confirmed as such, but we are at least aware of a near Jovian-analogue in planet-star separation and mass with MOA-2010-BLG-477L~b \citep{blaetal2021}. If that star is shown to be polluted, then we might be able to make qualitative statements about the role mean motion resonances are playing in that system's dynamical evolution.

The high computational expense of our simulations precludes a wider exploration of parameter space in this investigation. Amongst our ``wish list" of potential extensions to this investigation include considering an interior, rather than exterior, planet, and determining how to scale down the mass of the planet while scaling up the resolution in order to still be able to detect resonant signatures. \cite{veretal2021} did perform a wider parameter space, but at a much lower resolution.

Our results here also may aid investigations which derive eccentric-orbit stability boundary and timescale criteria \citep{georgakarakos2008,petetal2018}, as well as chaos-triggering and resonant overlap criteria \citep{quillen2011,muswya2012,decetal2013,hadlit2018,petetal2020,tametal2021,ratetal2022}, and determine how these criteria might change with mass shifts as the star evolves. 

What we have integrated here is many instances of the eccentric, non-coplanar restricted three-body problem with mass loss. However, as suggested by \cite{rametal2015}, this scenario is complex-enough to not necessarily be well-characterized by a single class of dynamical measures (e.g. Hill stability, Lagrange stability, angular momentum deficit, chaotic zone boundaries, resonance overlap). For example, none of the stable $4$:$3$ systems are Hill stable, and many of the resonant overlap investigations cited above assume circular orbits and multiple massive planets.

Indeed, one potential extension of this work is to include a second planet into our simulations. Then, secular resonances might become prominent, and the studies of \cite{smaetal2018} and \cite{smaetal2021} may better be used as a benchmark for comparison. Perhaps though the most physically important extension to these simulations would be to model the radiative effect on the test particles. Our results here hold only when the test particles are large enough ($\gtrsim 10^3$ km) to remain unaffected. Smaller debris would instead move about significantly during the simulations from stellar radiation alone \citep{bonwya2010,donetal2010,veretal2015a,veretal2019,zotver2020,feretal2022}.

\section{Summary}

We have produced high-resolution resonant portraits (Figs. \ref{FullWD}-\ref{FullStable}) of an evolved one-planet system with $10^5$ asteroids, and compared the $N$-body simulation results to a semi-analytical libration width calculator \citep{galetal2021} that does not rely on traditional disturbing function expansions and allows for full three-dimensional inputs. These portraits include regimes of both crossing and non-crossing orbits, and provide insight into the escape, pollution and stability efficiencies of different resonances {\rev at different stages of stellar evolution} (Table \ref{OutcomeTable}). Our maximum libration width computations are robust (e.g. Fig. \ref{Zoom31WDvarpi}) and usefully show (e.g. Fig. \ref{Zoom1}) that we can use main-sequence values to accurately determine the domain of influence of different resonances -- saving time and computational resources for future resonance-based investigations of evolved planetary systems.

\section*{Acknowledgements}

{\rev We thank Tabare Gallardo for providing valuable insight which has significantly improved the manuscript.} DV gratefully acknowledges the support of the STFC via an Ernest Rutherford Fellowship (grant ST/P003850/1). 

\section*{Data Availability}

The simulation inputs and results discussed in this paper are available upon reasonable request to the corresponding author.

\label{lastpage}
\end{document}